\documentstyle [12pt] {article}
\newcommand{\cs}[3]{{{#3} \brace {#1 #2}}}
\begin{document}
\title {Path and Path Deviation Equations in Kaluza-Klein Type Theories}
\maketitle
\begin{center}
{\bf {M.E.Kahil{\footnote{Mathematics Department, Modern Sciences and Arts University, Giza, EGYPT}} {\footnote{Mathematics Department, The American University in Cairo, Cairo, EGYPT\\
e.mail: kahil@aucegypt.edu}}}
}
\end{center}

\abstract{Path and path deviation equations for charged, spinning
and spinning charged objects in different versions of
 Kaluza-Klein (KK) theory using a modified Bazanski Lagrangian have been derived. The significance of motion in five
 dimensions, especially for a charged spinning object, has been examined. We have also extended the modified
  Bazanski approach to derive the path and path deviation equations of a test particle in a version of non-symmetric
   K-K theory.}
\section{Introduction}

In an attempt to unify gravity and electromagnetism Kaluza(1921)
introduced a fifth dimension to describe electromagnetism.
Klein(1926) added a stringent cylindrical condition , which keeps
the extra dimension compact[1]. Following the scheme of
compactification many theories have developed KK ideas and
extended the process of compactification to include higher
dimensions as a way to unify many fields[2]. However, Wesson et
al. [3] have considered unification of geometry with matter by
dropping the cylindrical condition, and introducing non-compact
theories of higher dimensions based geometrically on the Campbell
-Magaard theorem. This approach has been emerged into two classes
of non-compact theories: brane theories [4] and space-time -matter
theories [5].

From this perspective, path  and path deviation equations play a
vital part to interpret the behavior of any particle  describing
any of the above mentioned theories. These equations offer a way
to test the new physics coming from the introduction of extra
dimensions[6]. The behavior of test particles and extended objects
could be used for examining additional phenomena embedded in
higher dimensions. Accordingly, we present a study of path and
path deviation equations for  charged, spinning and spinning
charged objects using different theories of KK. The path and path
deviation equations could also be used to detect the cosmological
variation of spin, and to study the evolution of the angular
momentum of galaxies, pulsars and high energy primordial objects
[7] using a gyroscopic motion in 5-dimensions. KK theories have
been extended to include different types of non-symmetric theory
of gravity. One such trial has been done by Kalinowski to unify
gravity and gauge fields using a multidimensional manifold in the
Jordan-Thirry manner[8].

The aim of the present work is to extend the Bazanski approach[9]
into 5D in order to derive some versions of path and path
deviation equations
 in multidimensional space for different
 objects such as charged, spinning and spinning charged particles, with taking the status of extra dimension
as either compact or non-compact. Moreover, we are going to apply
 the Bazanski approach to derive the path and path deviation equations for a test particle moving in
 non-symmetric theories of gravitation in 5D .\\ The paper is organized into the following steps: \\
(1) Describing path equations and their corresponding path deviation equations in 4D. \\
(2) Extending these equations into 5D. \\
(3) Comparing and contrasting each equation in compact spaces with
the corresponding equations for non-compact spaces.
\section{Motion in 4D}
\subsection{Path \& Path Deviation Equations in Riemannian Geometry}
 Geodesic and geodesic deviation equations can be obtained simultaneously by applying the action principle on the
 Bazanski Lagrangian [9]:
\begin{equation}
L= g_{\alpha \beta} U^{\alpha} \frac{D \Psi^{\beta}}{Ds},
\end{equation}
 where $\frac{D}{Ds}$ is the covariant derivative. This can be done if one takes the variation with respect
to the deviation vector ${\Psi^{\rho}}$ in order to derive the
geodesic equation:
\begin{equation}
\frac{dU^{\alpha}}{ds} +\cs{\mu}{\nu}{\alpha}U^{\mu}U^{\nu}=0 ,
\end{equation}
where $\cs{\mu}{\nu}{\alpha}$ is the Christoffel symbol. If one
takes the variation with respect to the unit tangent vector
$U^{\rho}$,  one derives the geodesic deviation equation:
\begin{equation}
\frac{D^{2}\Psi^{\alpha}}{Ds^{2}} = R^{\alpha}_{. \beta \gamma
\delta}  U^{\beta}U^{\gamma} \Psi^{\delta}.
\end{equation} where $R^{\alpha}_{\beta \gamma \delta}$ is the Riemann- Christoffel curvature tensor.
 It is worth mentioning that the Bazanski approach has been successfully  applied  in  geometries different from the Riemannian  [10],[11].
Now, Lagrangian (1) can be amended to describe path and path deviation equations
of charged, spinning and spinning charged particles if we introduce the following Lagrangian:
\begin{equation}
L = g_{\alpha \beta} U^{\alpha}{\frac{D \Psi^{\beta}}{Ds}} + f_{\beta}\Psi^{\beta}
\end{equation}
such that
$${ f_{\beta} =  a_{1} F_{\alpha \beta} U^{\beta} + a_{2} R_{\alpha \beta \gamma \delta} S^{\gamma \delta} U^{\alpha}} ,$$
 where $a_{1}$ and $a_{2}$ are parameters that may take the values ${\frac{e}{m}}$ and ${\frac{1}{2m}}$ respectively,$F^{\mu}_{.\nu}$ is an Electromagnetic tensor and  $S^{\gamma \delta}$ is the spin tensor . These parameters have  to be adjusted with their counterparts in the original Lorentz force equation [12], the Papapetrou equation [13] and the Dixon equation [14].

Applying the Bazanski approach for obtaining path and path deviation equations on Lagrangian(4) we get:\\
(i)The Lorentz charged equation  (for $a_{1}= \frac {e}{m}$ and $a_{2}= 0$)
\begin{equation}
\frac{dU^{\alpha}}{ds} +\cs{\mu}{\nu}{\alpha}U^{\mu}U^{\nu}= \frac{e}{m}F^{\mu}_{. \nu} U^{\nu},
\end{equation}
and the charged deviation equation [15]:
\begin{equation}
\frac{D^{2}\Psi^{\alpha}}{Ds^{2}}= R^{\alpha}_{.\mu \nu\rho}U^{\mu}U^{\nu}\Psi^{\rho} +\frac{e}{m}(F^{\alpha}_{.\nu} \frac{D \Psi^{\nu}}{Ds}+F^{\alpha}_{.\nu ; \rho}U^{\nu}\Psi^{\rho}).
\end{equation}
(ii)The Papapetrou equation for spinning objects  ( for $a_{1}= 0$ and $ a_{2}= \frac {1}{m} $ )

\begin{equation}
\frac{dU^{\alpha}}{d s}+\cs{\mu}{\nu}{\alpha}U^{\mu}U^{\nu}= \frac{1}{2m}
R^{\alpha}_{. \mu \nu \rho} S^{ \nu \rho} U^{\mu}.
\end{equation}
The spinning deviation equation becomes [16]:
\begin{equation}
\frac{D^{2}\Psi^{\alpha}}{Ds^{2}}= R^{\alpha}_{.\mu \nu\rho}U^{\mu}U^{\nu}\Psi^{\rho} +\frac{1}{2m}( R^{\alpha}_{. \mu \nu \rho} S^{\nu \rho} \frac{D \Psi^{\mu}}{Ds}+
R^{\alpha}_{\mu \nu \lambda}S^{\nu \lambda}_{.; \rho}U^{\mu}\Psi^{\rho} + R^{\alpha}_{\mu \nu \lambda; \rho }S^{\nu \lambda} U^{\mu} \Psi^{\rho})
\end{equation}
(iii)The Dixon Equation for spinning charged objects(for $a_{1}= \frac{e}{m}$ and $a_{2}= \frac {1}{m}$ ) [14]:
\begin{equation}
\frac{dU^{\alpha}}{d s}+\cs{\mu}{\nu}{\alpha}U^{\mu}U^{\nu}=   \frac{e}{m}F^{\mu}_{. \nu} U^{\nu}+\frac{1}{2m}
R^{\alpha}_{. \mu \nu \rho} S^{ \nu \rho} U^{\mu},
\end{equation}
and its spinning charged deviation equation becomes:
$$
\frac{D^{2}\Psi^{\alpha}}{Ds^{2}}= R^{\alpha}_{.\mu \nu\rho}U^{\mu}U^{\nu}\Psi^{\rho} +\frac{e}{m}(F^{\alpha}_{.\nu} \frac{D \Psi^{\nu}}{Ds}+F^{\alpha}_{.\nu ; \rho}U^{\nu}\Psi^{\rho})+ \frac{1}{2m} R^{\alpha}_{.\mu \nu\rho}U^{\mu}U^{\nu}\Psi^{\rho}
$$
\begin{equation}
~~~~~~~~~~~~~~ + \frac{1}{2m}( R^{\alpha}_{. \mu \nu \rho} S^{\nu
\rho} \frac{D \Psi^{\mu}}{Ds}+ R^{\alpha}_{\mu \nu \lambda}S^{\nu
\lambda}_{.; \rho}U^{\mu}\Psi^{\rho} + R^{\alpha}_{\mu \nu
\lambda; \rho }S^{\nu \lambda} U^{\mu} \Psi^{\rho}) .
\end{equation}

Papapetrou [17] has derived an equation describing a spinning object which
is able to precess:

\begin{equation}
 \frac{D}{Ds}( m U^{\alpha} + U_{\rho}\frac{D S^{\alpha \rho}}{Ds})= \frac{1}{2}
R^{\alpha}_{. \mu \nu \rho} S^{\nu \rho } U^{\mu}  .
\end{equation}

Using the Bazanski approach, we can suggest the following Lagrangian:
\begin{equation}
L= g_{\alpha \beta} ( m U^{\alpha} + U_{\rho}\frac{D S^{\alpha \rho}}{Ds}) \frac{D \Psi^{\beta}}{Ds} + R_{\alpha \beta \gamma \delta} S^{\gamma \delta} U^{\beta} \Psi^{\alpha},
\end{equation}
which can be used to derive equation (11) and to obtain its corresponding deviation equation in the following way:
$$
\frac{D^{2}\Psi^{\alpha}}{Ds^{2}}=  R^{\alpha}_{.\mu \nu\rho}U^{\mu}( m U^{\nu} + U_{\beta}\frac{D S^{\nu \beta}}{Ds})\Psi^{\rho}+ g^{\alpha \sigma}g_{\nu \lambda}( m U^{\lambda} + U_{\beta}\frac{D S^{\lambda \beta}}{Ds})_{; \sigma} \frac{D \Psi^{\nu}}{Ds}
$$
\begin{equation}
~~~~~~~+ R^{\alpha}_{. \mu \nu \rho} S^{\nu \rho} \frac{D
\Psi^{\mu}}{Ds}+ R^{\alpha}_{\mu \nu \lambda}S^{\nu \lambda}_{.;
\rho}U^{\mu}\Psi^{\rho} + R^{\alpha}_{\mu \nu \lambda; \rho
}S^{\nu \lambda} U^{\mu} \Psi^{\rho} .
\end{equation}

\subsection{ Path \& Path Deviation Equations in Non-Symmetric Geometries} Path equations in one of the versions of non-symmetric geometries  e.g. Legar\'{e} and Moffat have been derived from  following Lagrangian [18]
\begin{equation}
L = g_{( \mu \nu)} U^{\mu}U^{\nu} + \lambda \hat{A}_ {\nu}U^{\nu},
\end{equation}
by taking the variation with respect to $U^{\sigma}$ to give
\begin{equation}
\frac{dU^{\alpha}}{ds} +\cs{\mu}{\nu}{\alpha}U^{\mu}U^{\nu}= \lambda g^{( \alpha \mu )} f_{[ \mu \nu ]} U^{\nu}
\end{equation}
where $g_{(\mu \nu)}$ is the symmetric part of the gravitational potential tensor,$\lambda$ is a parameter and, $f_{[ \mu \nu] }= \hat{A}_{\mu ,\nu} - \hat{A}_{\nu , \mu}$ is a skew symmetric tensor related to the Yukawa force.

Applying the Bazanski approach, we can derive (15) from the following Lagrangian:
 \begin{equation}
L= g_{(\alpha \beta )} U^{\alpha} \frac{D \Psi^{\beta}}{Ds}+
\lambda f_{\nu}\Psi^{\nu}.
\end{equation}
Using the same approach, we can show its corresponding deviation equation to be:
\begin{equation}
\frac{D^{2}\Psi^{\alpha}}{Ds^{2}}= R^{\alpha}_{.\mu \nu\rho}U^{\mu}U^{\nu}\Psi^{\rho} +\lambda(f_{\alpha_{.\nu}} \frac{D \Psi^{\nu}}{Ds}+f^{\alpha}_{.\nu ; \rho}U^{\nu}\Psi^{\rho}) .
\end{equation}

But, if we consider the following Lagrangian:
\begin{equation}
L = {\bf{g}}_{\mu \nu} U^{\mu} \frac{D \Psi^{\nu}}{D \tau} +
\lambda f_{\nu}\Psi^{\nu},
\end{equation}
where ${\bf{g}}_{\mu \nu} = g_{(\mu \nu)} + g_{[\mu \nu]} $, and follow the Bazanski approach to get the
path and path deviation equations related to this type of geometry by taking the variation with respect to $\Psi^{\sigma} $ and $U^{\sigma}$ respectively, we can obtain:
\begin{equation}
 \frac{dU^{\alpha}}{ds} +\cs{\mu}{\nu}{\alpha}U^{\mu}U^{\nu}= \lambda {\bf{g}}^{ \alpha \mu } f_{[ \mu \nu ]} U^{\nu} +
  {\bf{g}}^{\alpha \sigma } g_{[ \nu \sigma  ]; \rho} U^{\nu}U^{\rho} ,
\end{equation}
and the path deviation equation becomes:
\begin{equation}
\frac{D^{2}\Psi^{\alpha}}{Ds^{2}}= R^{\alpha}_{.\mu
\nu\rho}U^{\mu}U^{\nu}\Psi^{\rho}+2 {\bf{g}}^{\sigma \alpha} (
g_{[\nu [ \sigma];\rho] } ) \frac{D \Psi^{\nu}}{Ds}U^{\rho}
+\lambda(f^{\alpha}_{.\nu} \frac{D
\Psi^{\nu}}{Ds}+f^{\alpha}_{.\nu ; \rho}U^{\nu}\Psi^{\rho}).
\end{equation}
It is clear that the difference between (15 ) and (19) is  related to absence of the spin of the source in (15). Thus, from (19) it is possible
to find an interaction between the spin of the source and the skew field [19]. Kalonowski [20]
has extended Moffat's version [21] which is described in Einstin-Cartan geometry, to establish a relation between the mass and fermion current  curves space-time while the spin of the source is twisting it.

Moreover,  Wanas and Kahil have extended the Bazanski approach, applying it in Einstein non-symmetric geometries[22] to reach the conclusion that paths in these geometries are naturally quantized (in the Planck sense of quantization)[10]. This type of natural quantization of paths exists in absolute parallelism geometries as well [11].

\section{Motion in 5D}
The problem of motion in higher dimensions is an intriguing problem.  The significance of motion in higher dimensions may yield
some indications with regards to the principles that should be followed when describing motion in 4-dimensions, i.e an equation which governs the motion in
4-dimensions [5].  In the present work, we will examine the effect
of non gravitational forces on the current motion, i.e. should this motion be absorbed into the extra dimension or remain unchanged from
the usual equation of motion in 4-dimensional space apart from increasing  the dimensions?

\subsection{The Bazanski Approach in 5-Dimensions}
In an attempt to derive path and path deviation equations in 5-dimensions, we extend the Bazanski Lagrangian to 5D:
\begin{equation}
L = g_{_{AB}}{U^{A}} {\frac{D \Psi^{B}}{DS}}
\end{equation}
where $(A=1,2,3,4,5)$.
By taking the variation with respect to the deviation vector $\Psi^{C}$ and the tangent vector $U^{C}$, we obtain the geodesic and geodesic deviation equations respectively,
\begin{equation}
\frac{dU^{C}}{dS} + \cs{A}{B}{C} U^{A}U^{B}=0,
\end{equation}
and
\begin{equation}
\frac{D^{2}\Psi^{C}}{DS^{2}} = R^{C}_{_{. ABD}} \Psi^{D}
U^{A}U^{B}
\end{equation}

\subsection{Compact Spaces }
The process to unify electromagnetism (gauge fields) and gravity
depends on extra component(s) of the metric using the cylinder condition [1]. Some authors
believe that compact dimensions can be tested, for example, by examining
the rate of energy released as a result of gravitational waves from binary pulsars [23].

In our study,  we derive the same geodesic and geodesic deviation equations given by Kerner et al. [15] using the Bazanski approach:
\begin{equation}
\frac{dU^{\mu}}{dS}+ \cs{\nu}{\lambda}{\mu}U^{\nu}U^{\lambda} +(
\frac{dx^{5}}{dS} +A_{\nu} \frac{dx^{\nu}}{dS}) F_{.\lambda}^{.
\mu}U^{\lambda}=0,
\end{equation}

\begin{equation}
\frac{d}{dS}( \frac{dx^{5}}{dS}+ A_{\mu}\frac{dx^{\mu}}{dS} )=0.
\end{equation}
where $Q \equiv \frac{dx^{5}}{dS}+ A_{\mu}\frac{dx^{\mu}}{dS}$ is
constant along the 5D geodesics i.e.
 $$
\frac{q}{m}= \frac{dx^{5}}{dS}+ A_{\mu}\frac{dx^{\mu}}{dS}.
$$
Consequently, (24) becomes:
\begin{equation}
\frac{dU^{\mu}}{dS}+ \cs{\nu}{\lambda}{\mu}U^{\nu}U^{\lambda} + \frac{q}{m} F_{\lambda}^{. \mu} U^{\mu} =0,
\end{equation}
with $ds^{2} = (1-Q^{2})dS^{2}$
and its corresponding path deviation equation becomes:
\begin{equation}
\frac{D^{2} \Psi^{\mu}}{DS^{2}} = R^{\mu}_{. \rho \nu \lambda}U^{\rho}U^{\nu}\Psi^{\lambda}+ \frac{q}{m} ( F^{\mu}_{.\nu ; \rho} U^{\nu} \Psi^{\rho} + F^{\mu}_{.\nu}\frac{D\Psi^{\nu}}{DS} ) + F^{\mu}_{.\lambda}U^{\lambda}( \frac{d}{dS}(A_{\lambda}\Psi^{\lambda}+ \Psi^{5})+ F_{\nu \rho } U^{\nu}\Psi^{\rho})
\end{equation}
and
\begin{equation}
\frac{d}{dS}(( \Psi^{5}+ A_{\lambda} \Psi^{\lambda} ) + F_{\lambda \rho} U^{\rho}\Psi^{\lambda}) = 0
\end{equation}
A charged particle whose behavior is described by the Lorentz equation in 4D behaves as
a test particle moving on a geodesic in 5D. This result is obtained from the usual Basanski method  in 5D rather than its modified method in 4D.\\
\subsection{Non-Compact Spaces}
In an attempt to unify geometry and matter, Wesson and his
collaborators[3] have assumed that $g_{AB,5} \not= 0$, which is
applied in the brane world models[4] and space-time-matter
theories [5]. The idea of non-compact spaces
is based upon the Campbell-Maagard theorem [24]. Using this approach,  Wesson [25] has found that: \\
{(1)}Massive particles travelling on a time-like geodesic in 4-dim
can be regarded as traveling on a null-geodesic in 5D. This is
obvious
as an implication of the behavior of wave-like particles in a double slits experiment.\\
{(2)} Massive particles travelling on any path may exhibit changes their rest mass because there is a direct contact with the fifth force. In this case, the path equation will be a generalization of the problem of moving particles having variable mass in classical mechanics.

It is well known that the path equation of a charged object described in non-compact space [26] is given by
\begin{equation}
\frac{d U^{\alpha}}{dS}+ \cs{\mu}{\nu}{\alpha}U^{\mu} U^{\nu}= n F^{\alpha}_{\mu}U^{\nu} U^{\nu} + \epsilon n^{2} \frac{\Phi^{; \alpha}}{\Phi^{3}} - A^{\alpha} \frac{dn}{dS} - g^{\alpha \lambda}{\frac{dx^{5}}{dS}}(n A_{\lambda,5}+ g_{\lambda \mu ,5} \frac{dx^{\mu}}{dS})
\end{equation}
and
\begin{equation}
\frac{d}{dS}\epsilon \Phi^{2}( \frac{dx^{5}}{dS}+
A_{\mu}\frac{dx^{\mu}}{dS} )=0
\end{equation}
where $ n= \epsilon \Phi^{2}(\frac {dx^{5}}{dS}+ A_{\mu}U^{\mu})$,  $\Phi $ is a scalar potential, and $\epsilon = \pm 1$ depending on whether the extra dimension is space-like or time-like respectively.  This leads to  $ \frac{q}{m} = \epsilon \Phi^{2}(\frac {dx^{5}}{dS}+ A_{\mu}U^{\mu})$ in which its scalar field affects the ratio of charge to mass.

However, the above equation has two main defects:
{\underline{it is not gauge invariant}}, and
{\underline{the additional extra force from an extra dimension is parallel to the four vector velocity}} i.e. $ f_{\mu}U^{\mu} \not= 0.  $

Ponce de Leon [26] has dealt with these two defects by using various
types of transformations in order to make (29) and (30) like  the geodesic
equation in its usual form:
\begin{equation}
\frac{d^2 \xi^{A}}{dS^{2}}+ \cs{B}{C}{A} \frac{d \xi^{B}}{dS} \frac{d \xi^{C}}{dS} =0,
\end{equation}
where $\xi^{A}$ is the projected 5D velocity. This allow us to introduce its corresponding Bazanski Lagrangian:
\begin{equation}
L = g_{AB} \frac{d \xi^{A}}{dS} \frac{D \Psi^{B}}{DS}
\end{equation}
which gives its geodesic deviation equation as:
\begin{equation}
\frac{D^{2}\Psi^{A}}{DS^{2}}= 0
\end{equation}

\subsection{ Path \& Path Deviation Equations of Non-Symmetric Geometries in 5D}
We now consider the following Lagrangian:
\begin{equation}
L = {\bf{g}}_{_{AB}} U^{A} \frac{D \Psi^{B}}{D S} + \lambda
f_{_{C}}\Psi^{C},
\end{equation}

Applying the Bazanski approach to derive the path and path
deviation equations by taking the variation with respect to
$\Psi^{D} $ and $U^{D}$ respectively, we obtain:
\begin{equation}
 \frac{dU^{A}}{dS} +\cs{B}{C}{A}U^{B}U^{C}= \lambda {\bf{g}}^{ AD } f_{_{[ D C ]}} U^{C} + {\bf{g}}^{A D } g_{_{[ C D  ]; M}} U^{C}U^{M} ,
\end{equation}
 and
\begin{equation}
\frac{D^{2}\Psi^{A}}{DS^{2}}= R^{A}_{_{.B CD
}}U^{B}U^{C}\Psi^{D}+2 {\bf{g}}^{DA } ( g_{[A [ D];M] } ) \frac{D
\Psi^{C}}{DS}U^{M} +\lambda {\bf{g}}^{DA}(f_{A_{.C}} \frac{D
\Psi^{C}}{Ds}+f_{A C ; M}U^{C}\Psi^{M}).
\end{equation}
In one version of K-K Non-symmetric theory of
gravity, Kalnowski [8] has summarized the role of the extra
dimension in the following matter: mass and fermion current curve
the four dimensions, the spin of the source and the electromagnetic
potential twist the fifth dimension.
\section{Rotation in 5D}
The concept of rotation in higher dimensions is related to obtaining the governing equation of the current spinning object
 [27]. For a spinning gyroscope it is well known that the fifth equation is testing the rate of precession [28].
  Some authors believe that the study of two nearby free-falling gyroscopes could be used to examine the question of
   the existence of gravitational waves[16].

 We may apply the Bazanski approach on the following Lagrangian:

   \begin{equation}
   L=g_{_{AB}}U^{A}\frac{D\Psi^{B}}{DS} + \frac{1}{2m}
   R_{_{ABCD}}S^{CD}U^{B}\Psi^{A}
   \end{equation}
   to derive the path equation of a spinning object in 5D:
\begin{equation}
\frac{dU^{C}}{d S}+\cs{A}{B}{C}U^{A}U^{B}= \frac{1}{2m} R^{C}_{_{. A
B D  }} S^{B D} U^{A}
\end{equation}
The above equation describes spinning objects in compact spaces which satisfy the cylinder condition
 i.e $ g_{AB,5}=0$. This is identical to the Dixon equation if we project it into four dimensions. The fifth coordinate
 will contribute the electromagnetic tensor, which has already appeared in the Dixon equation.\\
Also, the original Papapetrou equation in 5D will be as follows:
\begin{equation}
 \frac{D}{DS}( m U^{A} + U_{E}\frac{D S^{AE}}{DS})= \frac{1}{2}
R^{A}_{_{. BCD}} S^{CD} U^{B}
\end{equation}
and its corresponding deviation equation will take the following form:
$$
\frac{D^{2}\Psi^{A}}{DS^{2}}=  R^{A}_{_{.B C D}}U^{B}( m U^{C} + U_{_{E}}\frac{D S^{C E}}{DS})\Psi^{D}+ g^{A C}g_{_{B E}}( m U^{E} + U_{_{O}}\frac{D S^{E O}}{DS})_{; C} \frac{D \Psi^{B}}{DS}
$$
\begin{equation}
~~~~~~~+ R^{A}_{. BCD} S^{CD} \frac{D
\Psi^{B}}{DS}+ R^{A}_{_{B C E}}S^{C E}_{.;
D}U^{B}\Psi^{D} + R^{A}_{_{B C E; D}
}S^{C E } U^{B} \Psi^{D} .
\end{equation}
These equations could be used to study the behavior of  spinning charged objects
 that exhibit precession  e.g. neutron stars,  compact objects..etc.

 In non-compact spaces with $R_{ABCD}=0$ , it is found that a spinning particle is moving on a geodesic in 5D rather than the Papapetrou equation [29]. This leads us to suggest that in non-compact spaces, satisfying the Campbell-Magaard theorem, spinning particles and spinning charged particles as well as test particles are moving along geodesics in 5D.
But if we consider the original Papapetrou equation in 5D, we can find out that it is different from the usual geodesic equation i.e.
\begin{equation}
 \frac{D}{Ds}( m V^{A} + V_{_{B}}\frac{D S^{AB}}{DS})= 0.
\end{equation}
On the contrary, its corresponding deviation equation is identical to
(33).
\section{Discussion and Conclusion}
 The Lagrangian required to derive the path and path deviation equations
  in higher dimensions  becomes the conventional Bazanski Lagrangian if
  the extra fields are described in the  fifth (higher) dimension. Path
  equations of different particles in 4D can be considered as the projection
  of the geodesic equation in 5D on 4D. But if the extra effect, non gravitational force,
  is not totally absorbed in the higher dimensional equations then the Bazanski Lagrangian
  should be amended like their counterparts in 4-dimension, which is seen in the 5D Papapertrou's
equation(38).

 It has been shown in this paper that the effect of
the compactness of the extra dimension(s) can be clearly perceived on moving objects. Also, we find that the effect of precession may distinguish between
tops and  test particles moving in a background whose 5D curvature
has vanished.

 In our study we have shown that the apparent
Papapetrou equation in 5D is merely the projection of the Dixon
equation in 4D. But if the space does not include electromagnetism
as an extra dimension, the Papapetrou equation in 4D remains the same
in 5D unless the extra dimension space is not compact satisfying
the Campbell-Magaard theorem. Also, as can be seen from (25) and (30) the path equations in compact and non-compact spaces display a contradictory aspect: the ratio of charge to mass
is constant in the case of compact space, while it is variable depending on the scalar field in case of non-compact spaces, which can be explored due to the effect of the cylinder condition on higher dimensions.

It was shown that in 4D, the Bazanski Lagrangian can be modified to describe path equations for charged, spinning  and spinning charged particles, as well as for spinning objects with precession and their corresponding path deviation equation. In an attempt to find the path and path deviation equations of the above mentioned particles in 5D, we have found that the Bazanski Lagrangian could remain unmodified if the non gravitational force could be absorbed into the higher dimension. Otherwise, the Bazanski Lagrangian must be amended.

In our study, we have also found that in non-compact spaces,  spinning objects and the non precessing ones are not following the same trajectory, although their path deviation equations are the same.

We have applied the Bazanski approach to determine path and path deviation equations of one version of the Einstein non-symmetric theories of gravity in 5D. This work could be extended to study the effect of compactness on the path equations in our future work.

In addition to our study, we would like to point out some trends based on extending  more than one time-like dimension.
 Recently, Chen [30] has shown that it is possible to interpret the double slit experiment
  using one than one time-like dimension. However, using  Wesson's approach [25], which is based on non-compact
 extra dimension, one could interpret the same effect. Also Chen [30] has obtained an equation of paths with different spins;
 these types of equations have also been obtained by Wanas [31] and applied by Wanas et al. to interpret the discrepancy of the COW-experiment [32] and to provide a consistent temporal model of SN1987A  using parameterized absolute parallelism geometry in 4D  [33]. So, introducing more than one time-like dimension to solve some problems in nature is a tempting suggestion that needs further discussion.

\section*{Acknowlgements}
The author would like to thank  Professors M.I. Wanas for his
guidance and remarks, M. Abdel-Megied  for his encouragement to
follow the present stream of research, G. De Young for his comments and
my colleagues, the members of the Egyptian Relativity Group . Also,
thanks to Professor K. Buchner for his kind
hospitality and invaluable comments during my stay in Munich in June
2004.

\section*{References}
 {[1]} Collins, P. Martin,A. and Squires, E. "(1989){\it{Particle Physics and Cosmology}}",John Wiley and Sons, New York. \\
{[2]} Gabaadaze, G.(2003)"Summer School on Astrophysics and Cosmology"
     17 June-5 July 2002. ICTP, Trieste, Italy \\
{[3]} Overduin, J.M. and Wesson, P.S. (1997), Physics Reports, {\bf{303}} \\
{[4]} Dick, R.(2001), Class. Quant. Grav., {\bf{18}}, R1. \\
{[5]} Ponce de Leon, J. (2003); gr-qc/ 0310780 \\
{[6]} Ponce de Leon, J. (2003) gr/qc0310078 \\
{[7]} Liko, T., Overduin, J.M. and Wesson, P.S. (2005) gr-qc/0311054. \\
{[8]} Kalinowski, M.W. (1989), {\it{Non-Symmetric Field Theory and its Applications}}, Institute of Theoretical Physics, Warsow University. \\
{[9]} Bazanski, S.I. (1989) J. Math. Phys., {\bf{30}}, 1018. \\
{[10]} Wanas, M.I. and Kahil, M.E.(1999) Gen. Rel. Grav.,
{\bf{31}}, 1921. ;
 gr-qc/9912007 \\
{[11]} Wanas, M.I., Melek, M. and Kahil, M.E.(1995) Astrophys.
Space Sci.,{\bf{228}}, 273.;
gr-qc/0207113. \\
{[12]} Sen, D.K. (1968), {\it{Fields and/or Particles}}, The Ryerson Press Toronto. \\
{[13]} Ravndal, F. (1980), Phys Rev. D, {\bf{20}}, 367. \\
{[14]} Dixon, W.G. (1970), Proc. Roy. Soc. Lond A{\bf{314}},499. \\
{[15]} Kerner,R. Martin,J. Mignemi,S. and van Holten,J-W. (2003), Phys Rev D, {\bf{63}}, 027502. \\
{[16]} Nieto, J.A., Saucedo, J. and Villanueva, V.M. (2003) Phys. Lett. {\bf{A312}}, 175 ; hep-th/0303123. \\
{[17]} Papapetrou, A.(1951), Proc. Roy.Soc. Lond A{\bf{208}},248. \\
{[18]} Legar\'{e}, J. and Moffat, J.W. (1996), Gen. Rel. Grav., {\bf{26}}, 1221.\\
 {[19]} Buchner, K. and Kahil, M.E. (2004) Private communication,

 TUM Munich, Germany. \\
{[20]} Kalinowski, M.W.(1982), Phys. Rev.{\bf D 26}, 3149. \\
{[21]} Moffat, J.W. (1979), Phys. Rev.{\bf{D 19}}, 3554. \\
{[22]} Einstein, A. (1955) {\it The Meaning of Relativity}, Princeton Univ. Press 

 New Jersey.\\
{[23]} Durra R., Kocian, P. (2004) Class. Quant. Grav., {\bf{21}}, 2127. \\
{[24]} Dahia, E., Monte,M. and Romaro,C. (2003),
Mod.Phys.Lett.{\bf{A18}},1773;gr-qc/0209013 \\
{[25]} Wesson, P. (2004) Gen.Rel. Grav., {\bf{36}}, 451. \\
{[26]} Ponce de Leon, J. (2002); gr-qc/ 0104008 \\
{[27]} Seahra, S. (2002); gr-qc/0204032\\
{[28]} Liu, H. and Mashhoon, B. (2000), Phys. Lett. A, {\bf{272}},26 ;

 gr-qc/0050079 \\
{[29]} Liu,H. and Wesson, P.S (1996), Class Quan. Grav., {{\bf{11}}, 1341 .\\
{[30]} Chen, X. (2005), gr-qc/05010341.\\
{[31]} Wanas, M.I.(1998) Astrophys. Space Sci.,
{\bf{258}}, 237 ;
 gr-qc/9904019. \\
{[32]} Wanas, M.I., Melek, M. and Kahil, M.E.
(2000) Grav. Cosmol.,
{\bf{6} }, 319. \\
{[33]} Wanas, M.I., Melek, M. and Kahil, M.E. (2002) Proc. MG IX,
part B, p.1100, Eds. V. Gurzadyan et al., World Scientific Singapore.\\

\end{document}